\documentclass[]{spie}  

\usepackage{amsmath,amsfonts,amssymb}
\usepackage{graphicx}
\usepackage[colorlinks=true, allcolors=blue]{hyperref}

\title{Space UV polarimeters}

\author[a]{Coralie Neiner}
\author[a]{Adrien Girardot}
\author[a]{Jean-Michel Reess}
\affil[a]{LESIA, Paris Observatory, PSL University, CNRS, Sorbonne University, Universit\'e Paris Cit\'e, 5 place Jules Janssen, Meudon, France}

\authorinfo{Further author information: coralie.neiner@obspm.fr}

\begin{document} 
\maketitle

\begin{abstract}
Several space missions are proposed or planned for the coming two decades dedicated or including mid- to high-resolution spectropolarimetry on a wide UV band. This includes the European instrument Pollux for the NASA HWO flagship mission, the NASA SMEX candidate Polstar, and the French nanosatellite demonstrator CASSTOR. We are developing UV polarimeters for these missions thanks to a R\&D program funded by CNES. For the mid- and near-UV, i.e. above ~120 nm, birefringent material (MgF2) can be used to produce a polarimeter. This is the baseline for Polstar, CASSTOR, and the MUV and NUV channels of Pollux. Prototypes have been built and tested with excellent results, and further tests are ongoing to fully characterize them. For the FUV channel of Pollux however, it is not possible to use this technology and we have instead studied a design based on mirrors only. We will present the various missions and instruments, their technical challenges, as well as the R\&D work performed on UV polarimeters and the proposed design solutions.  
\end{abstract}


\section{INTRODUCTION}
\label{sec:intro}  

Mid- to high-resolution spectropolarimeters on a wide UV band are proposed for several space missions or projects and will allow new breakthrough science \cite{neiner2023}. Such instrument has never been built before and thus requires developments and testing before flight. The 3 projects currently under study are:
\begin{itemize}
\item The Pollux spectropolarimeter for the HWO (Habitable Worlds Observatory) flagship mission at NASA. Pollux will cover a very wide range of wavelengths from the FUV at 100 m to the near-IR. This will be done with several dedicated spectropolarimeters. While the exact details are still under study and will be subjected to trade-offs as the concept advances \cite{muslimov2024}, the current design includes 4 channels covering the FUV from 100 to 120 nm, the MUV from 120 to 200 nm, the NUV from 200 to 400 nm, and the visible and near-IR from 400 nm to $\sim$1.5 $\mu$m. The resolving power decreases from $\sim$120000 in the FUV to $\sim$60000 in the near-IR. HWO is expected to launch in the early 2040's. It is currently undergoing a maturation phase (GOMaP) at the end of which (2029) the exact suite of instruments will be selected. Pollux onboard HWO will be able to address many science topics, from cosmic structures (interstellar, intergalactic and circumgalactic media, galaxies, cosmology) to stars (hot and cool stars, white dwarfs, novae, supernovae, gamma-ray bursts,...) and exoplanets (atmospheric characterization, atmospheric escape, star-planet interactions, surface and interior composition), as well as solar system planets. 
\item The CASSTOR mission is a scientific and technological demonstrator for MUV and NUV spectropolarimeters. It is a 16U nanosatellite developed in France, which will observe bright hot stars. This will be the first instrument to perform UV spectropolarimetric measurements of stars over a large UV band, pioneering the UV spectropolarimetric science. It will thus also allow us to better specify the required measurement precision for future missions such as HWO. CASSTOR will host a 12 cm telescope and its spectropolarimeter will cover the wavelength range from 134 to 291 nm with a resolving power of $\sim$13000. It has already undergone a Phase 0 study at the Paris Observatory and CNES and is about to start Phase A.
\item Polstar \cite{scowen2022} is a proposal for a SMEX mission to NASA, equipped with a 40 cm telescope and a spectropolarimeter covering 120 to 285 nm with a resolving power of $\sim$20000. It will study the role of rapid rotation in the evolution of massive stars and the galaxy. It will be submitted to NASA in June 2025 and, if successful, launched in 2031. 
\end{itemize}

\section{NUV and MUV polarimeters}
\subsection {Design}
For the MUV and NUV channels of Pollux for HWO, as well as for Polstar, the polarimeter will be based on temporal polarization modulation achieved with birefringent material. In this UV domain, MgF2 is the only material that is both transparent and birefringent and is thus the material we have to use. Because it is the only available material, we cannot combine several materials to achromatize the polarimeter over the whole wavelength range of interest, as is often done in the visible range where more materials are available. Instead, we achromatize the efficiency of extraction of the polarimetric information \cite{deltoroiniesta2000}.

We optimize the design of the polarimeter to reach an efficiency of extraction of the polarimetric information over the full wavelength range that best fits our requirements. If we want to extract the Stokes QUV parameters simultaneously, we need to aim for an extraction efficiency of 1/sqrt(3) $\sim$ 0.57. If only one Stokes parameter is of interest, we aim for an extraction efficiency of 1. 

To optimize the design and reach the required efficiency, we create a modulator composed of thin MgF2 plates and we adjust the thickness and the optical axis angle of each of the plate as well as the angles at which we rotate the full stack of plates. 

The optimal plates are usually extremely thin, which makes them impossible to built. We thus replace each plate by a pair of plates whose difference in thickness is equal to the optimal thickness we computed. This pair of plates is equivalent to the optimized too thin plate, but we can use whichever thickness is convenient as long as we control the thickness difference between the two plates in the pair. Since MgF2 absorbs light, it is of course preferable to keep the thickness as small as possible. We typically opt for $\sim$0.3 mm plates. 

If the wavelength range is not too wide, the optimal modulator  is usually composed of 2 pairs of plates, i.e. 4 plates. For a wider wavelength range, 3 pairs, i.e. 6 plates, may be necessary. 

The modulator needs to be rotated to at least 4 angle positions to make a full Stokes IQUV measurement, but using more angle positions allows redundancy and increases the precision of the polarimetric measurements \cite{sabatke2000}. For Polstar and CASSTOR, since we prefer to keep the total measurement duration as short as possible, we opted for the minimum 4 angle positions, while for Pollux we are planning for 6 positions. The angles can be forced to be equidistant or we can tailor them to optimized  values. In the former case, the demodulation of the observations will be easier, while in the latter case we can further slightly optimize the efficiency of extraction of the polarimetric information \cite{sabatke2000}.

The modulator is followed by a Wollaston prism, also made of MgF2. This analyzer splits the light beam into two orthogonal states of polarization. The two beams then enter an echelle spectrograph. 

\subsection{Bench tests}

Such a polarimeter in MgF2 was first tested on a dedicated bench in the visible domain \cite{pertenais2016} and led to excellent results. We then tested it in the UV domain on an existing large high-resolution ($\sim$150000) UV spectrograph available at the Paris Observatory. We showed that the polarimeter also works in the UV but we were not able to determine the polarimetric precision that we could achieve \cite{Legal2018}. This is due in part to the imprecision of the flexible photographic plates available for this spectrograph and in part to contamination of the optics since the spectrograph is not installed in a clean room. We thus elected to perform further tests in the UV domain on a dedicated bench in a clean room and limit as much as possible the contamination. The design of this bench is a modified version of the CASSTOR nanosatellite design, i.e. it is miniaturized. At the time of writing, most of the pieces of the bench have been purchased or built, but we are awaiting the delivery of the optics of the collimator and objective to start the new experiment. We plan to carry on the tests in 2025. 

\subsection{Optical contacting and thermal tests}

Since the modulator is composed of several very thin plates of MgF2, flux is lost in Fresnel reflections and the modulator acts as a Fabry-Perot which creates unwanted polarized spectral fringes. To improve this, we put the plates in optical contact, to form a solid block of MgF2. Since MgF2 is birefringent, its coefficient of thermal expansion (CTE) depends on direction wrt the optical axis of the crystal. And since the optical axis of each plate differs, each plate will preferentially expand in a different direction. This produces stresses on the optical contact. 

We performed thermal tests to check how well the optical contact can stand thermal variations in space conditions\cite{legal2020}. We found that within a temperature range of -10 to +55 degrees Celsius, all our samples resisted thermal cycling. Below this temperature range however, the optical contact of some of our sample modulators got loose. None of the plates broke, even at low temperature, which means that the modulator would still be operational even in these conditions, but the demodulation matrix of the system would have to be revised if a contact breaks and such a recalibration is not easily done once in flight. Therefore we advice to maintain the temperature of the system between -10 and +55 degrees Celsius. 

\subsection{The CASSTOR demonstrator}

To further tests the polarimeter in real conditions and acquire the very first UV spectropolarimetric data of stars, we developed the CASSTOR nanosatellite project. The polarimeter is composed of a modulator with 2 pairs of MgF2 plates, rotating at 4 angle positions, followed by a MgF2 Wollaston prism. 

Because CASSTOR is a nanosatellite, its platform is not very stable, which is an important issue for spectropolarimetry. Indeed, while we acquire a full sequence of 4 modulator angles, the spectrum should not move on the detector. Otherwise the movement of the spectrum forces us to correct for flat-field differences between pixels and such a correction introduces errors that are larger than the polarimetric signature we are trying to detect in hot stars. For CASSTOR we computed that the spectrum should remain stable within 2.3 $\mu m$ (0.23 pixel) on the detector plane during the polarimetric sequence and we set a maximum sequence duration of 32 minutes. To achieve this stability, CASSTOR includes a Fine Guiding System with a tip-tilt mirror within the instrument, coupled to the attitude control system of the platform. 

\section{FUV polarimeters}
\subsection{Design}

For the FUV channel of Pollux (100-120 nm), the design presented above cannot be used. Instead we designed a polarimeter made of mirrors only. Indeed, any reflection of light on a mirror produces polarization of the light. By combining several mirrors at various angles and rotating the stack of mirrors as a whole, we can again produce temporal modulation of the polarization. A last mirror is added for the analyzer task. In this case however, only one beam is obtained at the output of the analyzer and thus 50\% of the signal is lost by default in the system. 

To optimize the modulator, we combine at least 3 mirrors and optimize their material or coatings and their respective angle. We can then also optimize the angles of modulation, as for the design above.

For Pollux, first attempts \cite{legal2020ApOpt} showed that the choice of mirror material is crucial but the complex optical indices of materials in the FUV are badly known. We proposed a first solution with a modulator in B4C and SiC, followed by an analyzer in taC \cite{legal2020ApOpt}. Since then, we have refined the design and now propose a more efficient solution with a modulator entirely in SiC, and an analyzer made of a SiO2 substrate coated with thin films of B4C and MgF2 \cite{girardot2024}. 

\subsection{Upcoming bench tests}

We will use the bench currently under construction for the NUV/MUV polarimeter to also test the FUV polarimeter. Most of the bench parts will be reused but the lamp will be replaced by a FUV lamp and the gratings and objective mirror of the spectrograph will be replaced to adapt to the wavelength range. We plan to carry on these tests once the NUV/MUV experiment is finished, i.e. we will likely proceed with this bench in 2026. 

\section{Conclusions}

Over the last years we have developed, optimized, and tested polarimeters covering a wide wavelength range in the NUV and MUV domains ($>$ 120 nm). Such polarimeters are ready for future space missions, even though we continue to characterize their precision and increase their technological readiness level (TRL). The launch of the CASSTOR nanosatellite demonstrator in particular will bring such instruments to a very high TRL. 

The development of polarimeters for the FUV domain, i.e. below 120 nm where MgF2 cannot be used anymore, has also started but is still at low TRL. It is needed in particular for the FUV channel of Pollux for HWO. We propose a design for this FUV polarimeter but more studies are needed to completely characterize it and test it. We plan to achieve this thanks to the test bench we are currently building at the Paris Observatory.

\acknowledgments 
We thank CNES for its continuous support and R\&T funding since 2012 for the UV polarimeter developments. We are also grateful to Juan Larruquert for useful discussions about FUV materials. 
 
\bibliography{ICSO2024_Neiner} 
\bibliographystyle{spiebib} 

\end{document}